\documentclass[a4paper,twocolumn,showpacs,amsmath,pra,amssymb]{revtex4-1} 
\usepackage[colorlinks=true,urlcolor=blue,citecolor=blue,linkcolor=blue]{hyperref}
\usepackage[T1]{fontenc}
\usepackage[latin9]{inputenc}
\usepackage{times}
\usepackage{color}
\usepackage{xspace}
\usepackage{amssymb,amsmath}
\usepackage{amsbsy}
\usepackage{graphicx}
\usepackage{epstopdf}
\usepackage{float}
\usepackage[T1]{fontenc}
\usepackage{pslatex}
\usepackage{contour}
\usepackage{color}
\usepackage[usenames,dvipsnames]{xcolor}

\usepackage[colorlinks=true,urlcolor=blue,citecolor=blue,linkcolor=blue]{hyperref}
\usepackage{color}
\definecolor{blaa}{RGB}{153,153,255}
\definecolor{filtered}{RGB}{153,0,51}

\usepackage{amsmath}
\usepackage{amssymb}
\usepackage{epstopdf}
\usepackage[T1]{fontenc}
\usepackage{times}

\newcommand{\fref}[1]{Fig.~\ref{#1}}

\newcommand{\Fref}[1]{Figure~\ref{#1}}

\newcommand{\Head}[1]{\paragraph*{#1}}

\definecolor{raw}{RGB}{255,177,100}
\definecolor{filtered}{RGB}{153,0,51}
\begin{document}
\title{Optical runaway evaporation for multi-BEC production}
\author{Amita B. Deb}
\author{Thomas McKellar}
\author{Niels Kj{\ae}rgaard}\email{nk@otago.ac.nz}
\affiliation{Jack Dodd Centre for Quantum Technology, Department of Physics, University of Otago, Dunedin, New Zealand.}
\date{\today}
\begin{abstract}
We report on parallel production of Bose-Einstein condensates (BECs) in steerable, multi-plexed crossed
optical dipole traps. Using a conventional trap-weakening evaporation scheme, where the optical trapping power is lowered,
we obtain an array of up to four independent BECs.
To improve evaporation efficiency, we propose to target each crossed trap site with a narrow auxiliary laser beam, creating an escape channel for energetic atoms. We experimentally demonstrate runaway evaporation using this scheme, which is characterized by very modest weakening in atomic confinement such that high densities and elastic collision rates can be maintained. Based on discretely time-shared optical tweezers, our approach is particularly suited for addressing the problem of simultaneosly cooling atoms in multiple traps clouds, providing the freedom to act locally and in a tailored fashion at individual trap sites.
\end{abstract}
\pacs{37.10.Gh, 67.85.Hj, 37.10.De, 64.70.fm}
\maketitle
\Head{Introduction}
Over the last century, atomic physics has undergone a remarkable evolution, from
beginnings in spectroscopy to advanced technologies harnessing exquisitely controlled interactions within atom-atom and atom-light systems, underpinning applications such as quantum simulations and quantum metrology. This has motivated the development of highly engineered methods to trap, cool, and manipulate atoms. While the first quantum degenerate Bose \cite{Anderson1995} and Fermi \cite{DeMarco1999} gases were obtained in traps based on magnetic fields, most contemporary experiments will, in their final stages, employ a configuration of laser beams for atom confinement via the optical dipole force. Since the seminal demonstration \cite{Barrett2001}, there has been a significant ongoing interest for all-optical implementations of BEC production, which offer simplicity and versatility \cite{Clement2009,Wilkowski2010,Jacob2011,Arnold2011,Lauber2011,Bruce2011,Marchant2012,Farkas2013,Jiang2013,Olson2013}.

As discussed in \cite{Jacob2011}, all-optical approaches for ultracold atoms face two issues. First, the spatial extent and the depth of the optical potential used for atom confinement should ideally match the size, density, and temperature of the atomic source. Second, to further lower the temperature of atoms captured, an efficient evaporative cooling scheme is important. In its simplest realization, forced evaporative cooling of an atomic ensemble confined by a configuration of laser beams will proceed when gradually lowering the optical power. The lowering of power reduces the potential depth at a trapping site and allows the most  energetic atoms to escape. It, however, also reduces the confinement of the trap and hence the density of atoms, thereby increasing the time it takes a sample to thermalize \cite{Adams1995}. This contrasts the accelerating rethermalization encountered during forced evaporative cooling in tight magnetic traps \cite{Ketterle1999b}. The key to reaching this highly desirable ``runaway" regime \cite{Ketterle1996} during evaporative cooling in optical dipole traps is a mechanism by which one can continually lower the truncation energy of the trap without compromising the trapping volume and thereby the elastic collision rate between atoms. In \cite{Hung2008} an external magnetic field gradient was introduced to ``tilt'' the potential of a crossed beam trap and in \cite{Clement2009} an auxiliary  large-diameter laser beam, offset with respect to the center of a tight single beam trap, was ramped up in power to reduce the trap depth. While both these schemes were successful in demonstrating runaway evaporation of a single dipole trapped sample they do not, however, lend themselves straightforwardly to multiple clouds located in more elaborate potentials. The perspective of multi-BEC production \cite{Onofrio2000,Boyer2006,Henderson2009} would, for example, seem attractive as a resource for feeding cold-atom scanning probe microscopes \cite{Gierling2011} at an increased rate, for BEC magnetic gradiometry \cite{Wood2014}, or as a means of replenishing atom lasers \cite{Robins2008}.
\begin{figure}
  \centering
  \includegraphics[width=\linewidth]{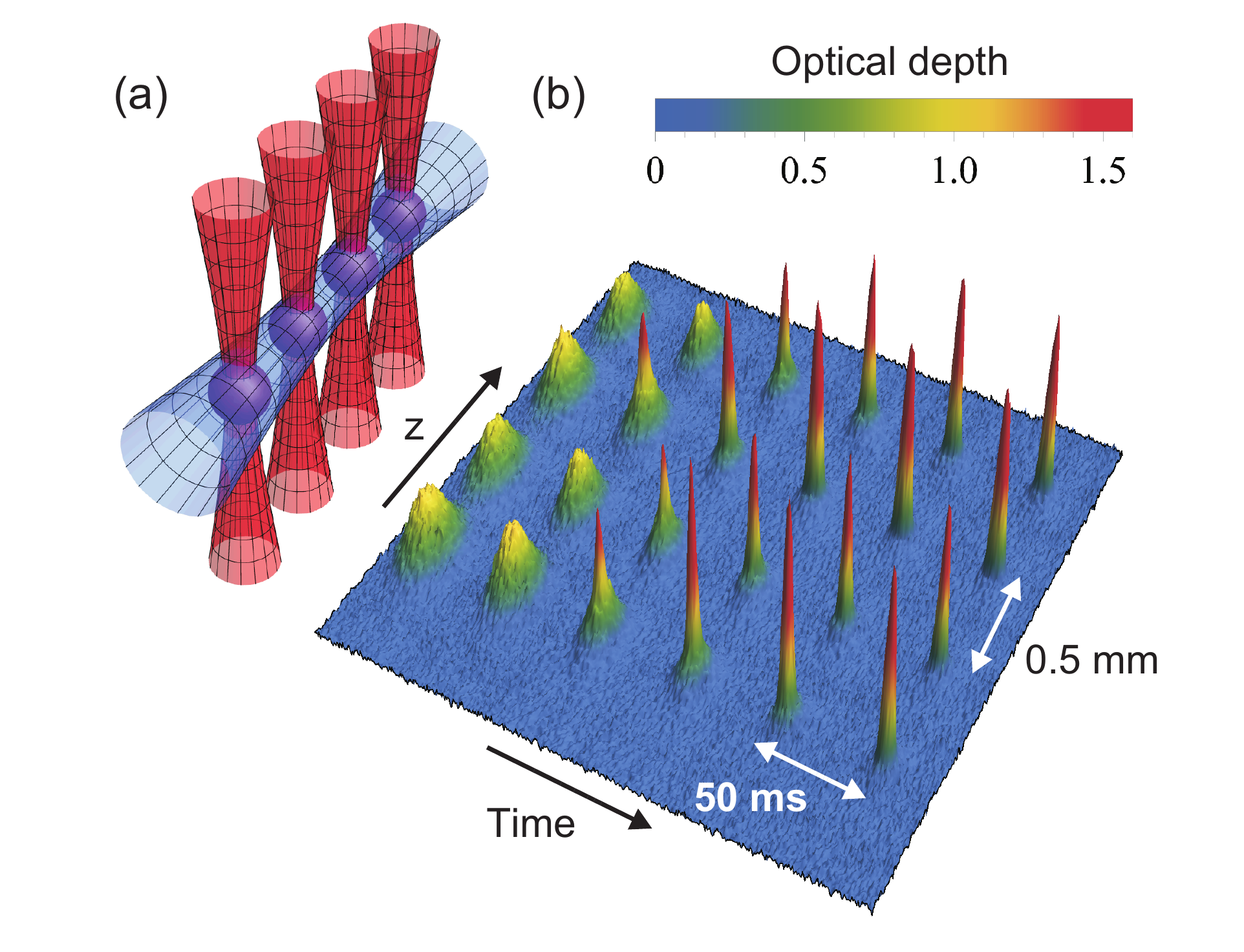}\\
  \caption{(Color online) (a) Multiplexed crossed-beam dipole traps; four-site case. Trap weakening evaporation proceeds by lowering the power of the horizontal beam. (b) Emergence and growth of four individual BECs during the final stage of lowering the horizontal beam power. }\label{Scheme 1}
\end{figure}

In this Rapid Communication, we consider forced evaporation at sites of multiplexed crossed dipole traps along a horizontal guide beam for atoms.  Using time-shared optical tweezers, we recently demonstrated the ability to split a thermal, bosonic $\rm^{87}Rb$ cloud containing $\sim3\times10^6$ atoms at a temperature of 1~$\rm\mu K$ into an array of 32 daughter clouds which could be spatially manipulated on a centimeter scale \cite{Roberts2014}. The question naturally arises if such arrays can be converted into BECs by means of on-site evaporative cooling? We find that using a trap weakening scheme for evaporation, where the horizontal guide beam power is progressively lowered, we can, indeed, produce up to four individual BECs. A central limiting factor for this basic evaporation scheme is, however, the aforementioned decrease in the rate of elastic collisions between trapped atoms, resulting from weakening their confinement. We propose and demonstrate an alternative evaporation scheme where a strong ghost tweezer beam in the proximity of a trapping site ``siphons'' out the hottest atoms. This does not happen at the expense of any appreciable weakening in the confinement at a trap site and we are able to establish runaway evaporation for this scheme.

\begin{figure}[b!]
  \centering
  \includegraphics[width=\linewidth]{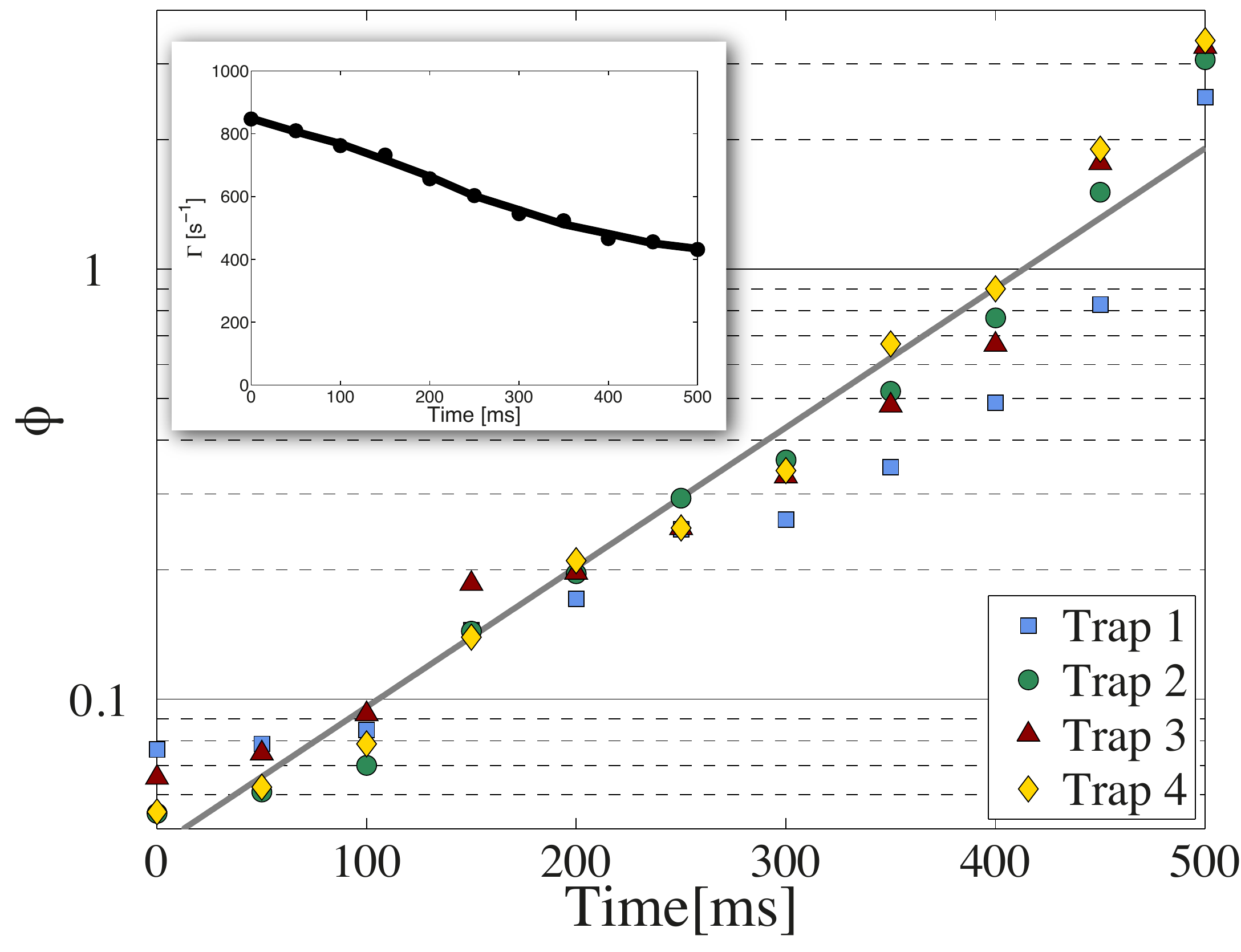}\\
  \caption{(Color online) Evolution of peak phase space densities during trap weakening evaporation; four-site case. Inset shows the corresponding elastic collision rate averaged over the four trap sites.}\label{Scheme1phase}
\end{figure}
\Head{Experimental setup}
We initially load an ultracold atomic cloud into a crossed optical dipole trap formed by a static horizonal ($z$-direction) laser beam (spot size radius of 60 $\mu$m) and a steerable, vertical ($y$-direction) tweezer beam  (spot size radius of 40 $\mu$m). The beams in the two dimensions are derived from the same 50~W 1064~nm single-frequency fiber laser (IPG Photonics,
YLR-50-1064-LP-SF) with optical powers being individually adjustable via separate acousto-optic modulators. Light for dipole trapping is delivered to the experiment in two single mode polarization maintaining fibers. The position of the dynamic tweezer can be set anywhere within a 6~mm~x~6~mm region in the $(x,z)$-plane by means of a 2D acousto-optic deflector (AOD). The details of this steerable optical tweezer system are described in \cite{Roberts2014}.  We load the atoms into the crossed beams from a Ioffe-Prichard (IP) magnetic trap. To achieve good spatial mode matching with this atom source, the vertical tweezer beam toggles rapidly at a rate of 200~kHz between two $z$-positions at $\pm12.5~\mu$m about the IP trap center; the $x$-position remains constant. This gives rise to a time averaged optical potential \cite{Molloy1998,Onofrio2000,Schnelle2008,Henderson2009,Zimmermann2011} which is elongated in the $z$-direction, matching the atomic source emittance for horizontal and vertical trap beam powers of 900~mW and 60~mW, respectively. In this way we obtain $4 \times 10^6$ atoms in the optical crossed trap with a temperature of $\sim$ 3 $\mu$K, from an initial sample of 5 $\times 10^6$ atoms at $\sim$ 2.5 $\mu$K.

\Head{Trap weakening evaporation: BEC quadruplets}
To investigate the prospects of producing multiple BECs in parallel, we ramp up the vertical trap beam power to 120~mW and split the initial (nearly-overlapped) double well into two separated traps moving in opposite directions along the horizontal guide beam to a distance of $\pm$ 0.5~mm from the center in 70~ms using a minimum jerk cost trajectory \cite{FLASH1985}. After a hold time of 10~ms, each of the resulting clouds are further split into two daughter clouds $\pm$ 0.25~mm from their initial locations in 40~ms. This results in four cold atomic clouds along the $z$-axis, each at a distance of 0.5~mm from its nearest neighbour as illustrated in \fref{Scheme 1}(a). At the end of the splitting phase, the atoms are held in their final locations for 100~ms while the vertical tweezer beam power is ramped down to $\sim90$~mW.  At this point we measure a mean atom number per well of $\bar{N}=7.0\times10^5$ with fluctuations $\lesssim6\%$ across sites; the four clouds are characterized by a temperature $T\sim  2.3~\mu$K, density $n \sim 3.4\times 10^{13}$ cm$^{-3}$, and peak phase space density $ \phi \,\sim \, 6.5\times 10^{-2}$.

Following the loading and splitting phases, we enforce evaporative cooling on the four clouds by reducing the horizontal laser beam power linearly in time from 900 mW to 310 mW over a period of 500 ms. This causes the trap to weaken and the effective depth in the vertical direction (with gravity taken into account) to decrease from 15 $\mu$K to 750~nK. In each well, the elastic collision rate between atoms at the beginning is very high ($\rm> 800~s^{-1}$) while the density is only moderate such that the inelastic three-body loss rate is small (0.25 s$^{-1}$). To characterize the progression of evaporation we acquire a series of absorption images of the quadruplet clouds in free fall following an 18~ms time-of-flight period initiated by switching off all trapping light during the ramp. From the absorption images we infer $N$, $T$ and $\phi$ for each well. \Fref{Scheme1phase} presents the evolution in $\phi$ over the four wells during evaporation. An even and exponential growth in $\phi$ can be observed and from our data we derive a characteristic evaporation efficiency parameter \cite{Ketterle1999b} $\gamma = -\frac{\ln(\phi/\phi_0)}{\ln(N/N_0)} \simeq 1.88$. A further reduction in the horizontal beam power from 310 mW to 290 mW in another 200~ms results in four nearly pure condensates of $N = \{4.0,4.5,4.5,4.8\}\times 10^4$. \Fref{Scheme 1}(b) shows a time sequence of absorption images (18~ms time-of-flight) of the quadruplets evolving into BECs.

\begin{figure}
  \centering
  \includegraphics[width=\columnwidth]{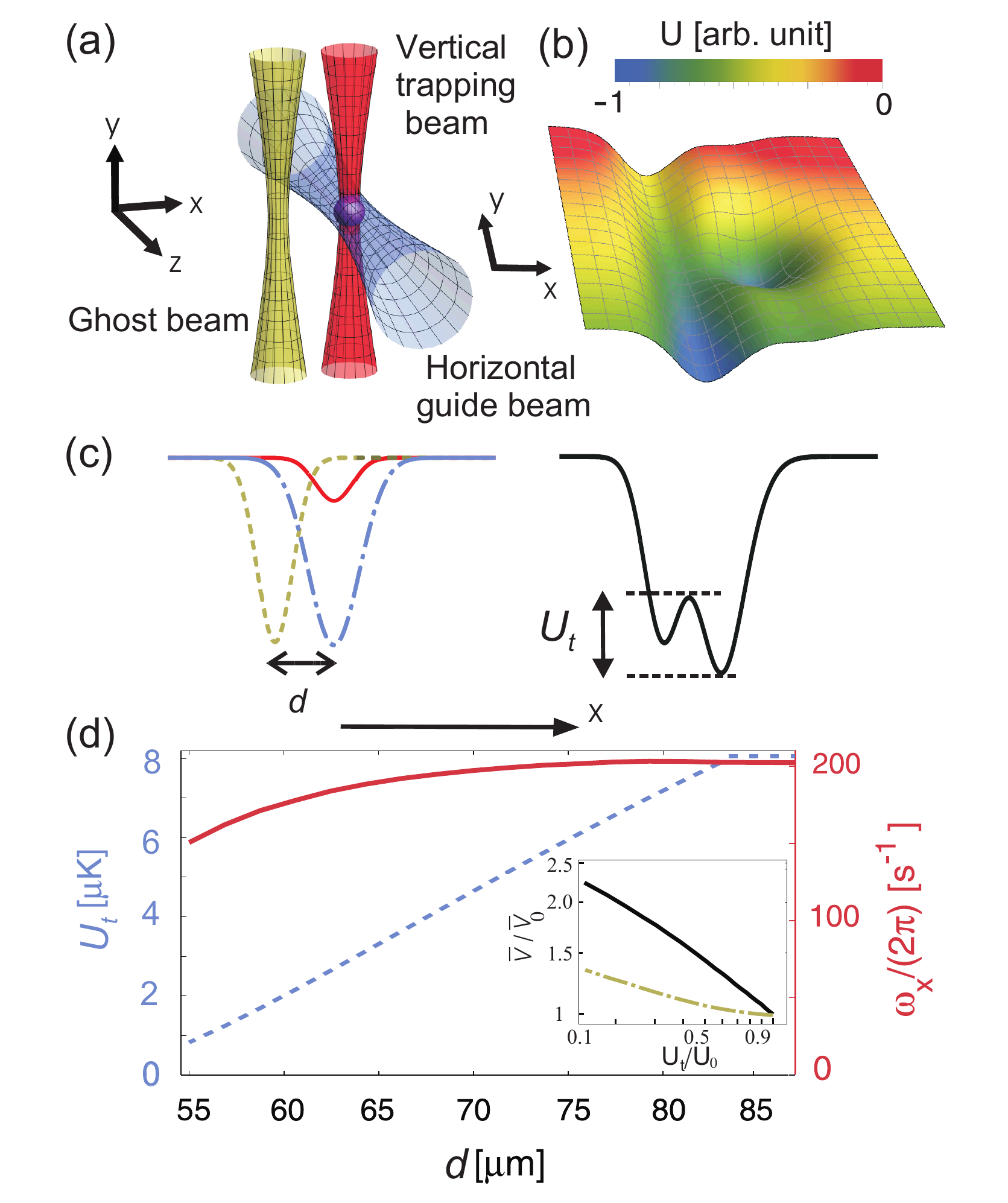}\\
  \caption{(Color online) Ghost beam evaporation. (a) Laser beam configuration:  Atoms are confined at the crossing of a horizontal and a vertical beam while a vertical ghost beam introduces an escape channel for hot atoms. (b) Surface plot showing the gravity-inclusive potential energy $U$ in the $(x,y)$-plane. (c)  Contributions from the horizontal beam (dash-dotted line), vertical trap beam (solid line) and vertical ghost beam (dashed line) to the trapping potential along the $x$-axis. Resultant potential is shown to the right with $U_t$ indicating the truncation energy. (d) Truncation energy $U_t$ and trapping frequency in $x$-direction $\omega_x$ as a function of ghost beam displacement $d$ along $x$-axis. Inset shows normalized scaling of trapping volume with respect to $U_t$ for ghost beam (dash-dotted line) and trap weakening (solid line) evaporation schemes.}\label{Scheme 2}
\end{figure}

 Using the atom numbers and temperatures obtained from the absorption images along with knowledge of the trap parameters and atomic scattering length, the elastic collision rate $\Gamma$ can be calculated \footnote{$\Gamma = n\sigma v_m$, where $n$ is the atomic density, $\sigma = 8\pi a_s^2$ is the s-wave scattering cross-section and $v_m = (16k_B T/\pi m)^{1/2}$ is the mean atomic velocity. Here $a_s$ is the s-wave scattering length and $m$ is the atomic mass.}. The inset of \fref{Scheme1phase} shows the evolution in $\Gamma$ for our four-well evaporation scheme. A significant $\sim 50\%$ reduction over the course of the ramp can be seen, and is the result of using a trap weakening evaporation scheme which relaxes the vertical confinement of atoms. While successful in attaining four individual BECs, the consequent drop in the thermalization rate renders efficient evaporation in this scheme challenging when splitting our initial atomic sample into more than four daughter clouds. For example, extending to the eight-well case, where the initial elastic collision rate at a trapping site is halved, we observed an insignificant gain of 2.5 in the peak phase space density for an order of magnitude loss in atom number, which was insufficient to reach degeneracy. Here, the atomic cloud takes longer and longer to thermalize as the trap is being weakened, while loss and heating mechanisms such as inelastic collisions, spontaneous photon scattering and collisions with background particles begin to play an important role. The situation can accurately be described as ``run out'' evaporation.
Evidently, by doubling the initial number of source atoms twice as many sites could be fed above the run out threshold number and eight BECs could be produced. Alternatively, the number of achievable BECs could be scaled up by employing a more efficient evaporation scheme and in the following we shall concern ourselves with the experimental demonstration of an optical runaway evaporation scheme for parallel multi-BEC production.

\Head{Runaway ghost beam evaporation}
\begin{figure}[tb!]
  \centering
  \includegraphics[width=\columnwidth]{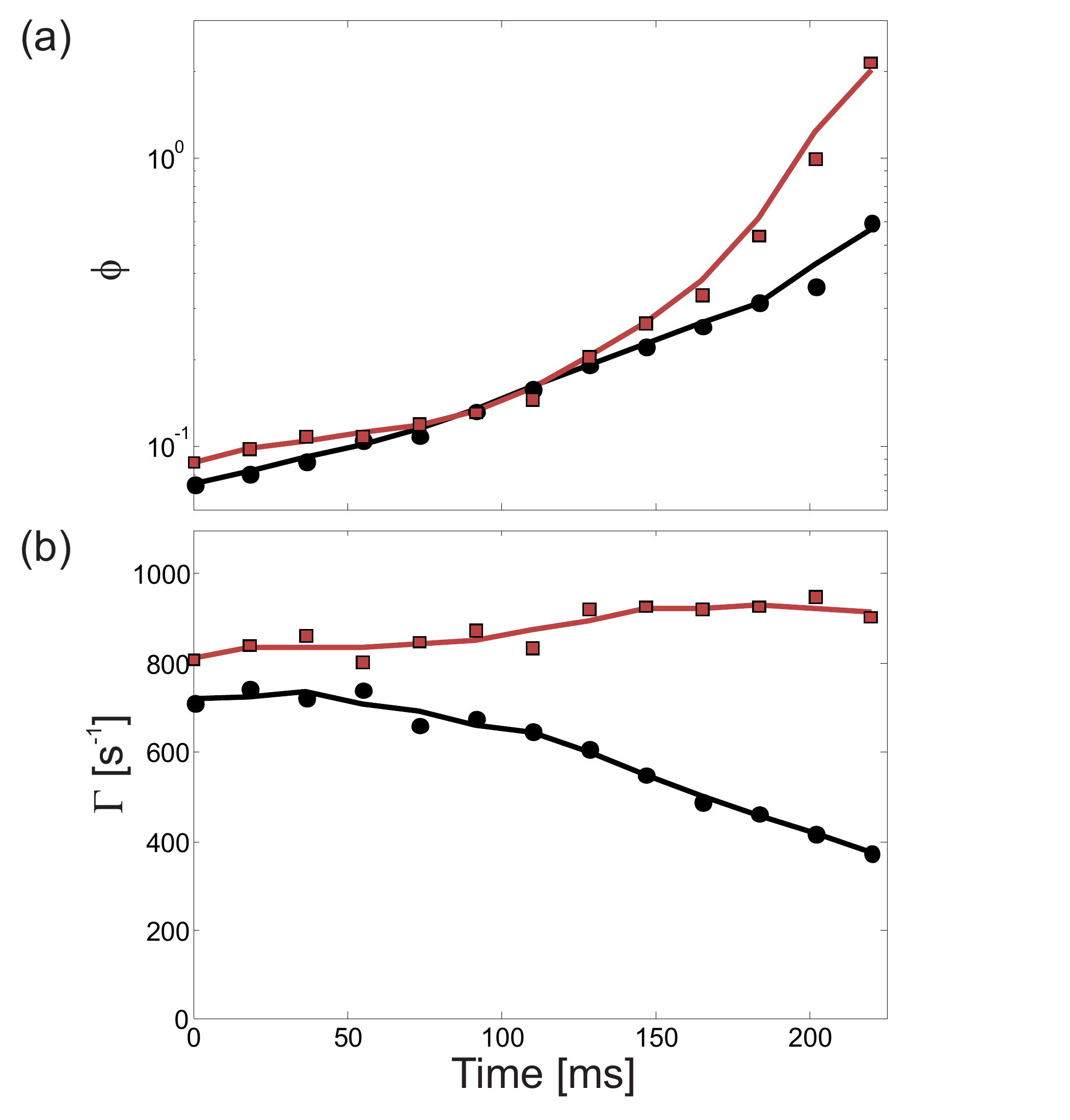}\\
  \caption{(Color online) Twin cloud evaporative cooling. Time evolution of (a) peak phase space density $\phi$ and (b) elastic collision rate $\Gamma$ during during ghost beam evaporation (red squares) and trap weakening evaporation (black circles). Values for $\phi$ and $\Gamma$ are two-cloud averages; lines present running means to guide the eye.}\label{Runaway}
\end{figure}
 As mentioned above, the $(x,z)$-coordinates for the dynamic vertical trapping beam are controlled by a 2D scanning device. For our BEC quadruplets in \fref{Scheme 1}(b) produced using trap weakening evaporation, $x$ was held constant while $z$ toggled rapidly between four values. The $(x,z)$-coordinates are encoded in the frequencies of harmonic signals driving two AODs with crossed deflection axes. By using a two-tone drive for the $x$-AOD, we can establish a situation where the vertical beam for each crossed trap along the $z$-axis is replicated in a ghost beam at a displaced $x$-position [see \fref{Scheme 2}(a)].  As illustrated in \fref{Scheme 2}(b,c) the presence of a ghost beam in the proximity of a crossed trap site can modify the potential energy landscape to create an escape channel for hot atoms above a certain truncation energy $U_t$: energetic atoms climb the barrier towards the ghost beam along which they funnel down under the action of gravity. A key requirement for establishing this situation is a peak intensity of the ghost beam comparable to that of the static horizontal beam. In
 \fref{Scheme 2}, for example, the horizontal beam, ghost beam, and vertical trap beam are assumed to have optical powers 520~mW, 290~mW and 90~mW, respectively. In \fref{Scheme 2}(d) we plot calculated values of $U_t$ and the $x$-direction trapping frequency as a function of ghost beam displacement $d$ from the trapping site. From this we infer the dependence of the mean trapping volume $\bar{V} \propto (\omega_x\omega_y\omega_z)^{-1}$ on the truncation energy which is shown in the inset of \fref{Scheme 2}(d). For comparison, the inset also presents the trap weakening case for which $\bar{V}$ blows up significantly faster with decreasing $U_t$ ---  clearly, the ghost tweezer scheme gives rise to a more favorable setting.

We investigate our proposed ghost beam evaporation scheme experimentally for a two-cloud case. An atom cloud is split into two daughter clouds which we position 1~mm apart, in a manner similar to before. At this point, we have 520~mW of optical power in the horizontal guide beam and 90~mW in the vertical tweezer beam. Each of the sites contains $\sim$ $1.6\times10^6$ atoms at 2 $\mu$K with a peak density of $3.6\times10^{13}$~cm$^{-3}$ and peak phase space density of $8\times 10^{-2}$. Ghost beams with an optical power of  290~mW are now introduced at a distance $d = 85$~$\mu$m from each crossed trap. The ghost beams are moved towards the waveguide linearly in 240~ms to a final position $d = 55$~$\mu$m corresponding to a truncation energy of 870~nK. The evolution of the two clouds during the ramp is found to be nearly identical, and \fref{Runaway} shows the mean values of $\phi$ and $\Gamma$ as a function of time for the first 220~ms. A faster than exponential growth in $\phi$ with time is evident, as is a slight increase in the elastic collision rate, both clearly indicating that the runaway regime has been reached. Onset of quantum degeneracy in each well is observed around t~=~220~ms at which point we have $\sim 2.4\times 10^5$ atoms in each trap. The evaporation process is highly efficient with $\gamma \sim 2.9$. Following a further 20~ms of evaporation, we are left with two nearly pure BECs with $N \sim 1.4 \times 10^5$. To benchmark our ghost beam scheme, \fref{Runaway} also presents the development in $\phi$ and $\Gamma$ under optimized trap weakening evaporation from the same initial condition. Here, while an exponential phase density increase is observed, the elastic collision rate significantly decreases with time, characteristic of trap weakening evaporation.

\Head{Discussion and conclusion}
As mentioned previously, our ghost beam evaporation scheme relies on ghost beam tweezers that are similar in strength to that of the horizontal guide beam. In the series of experiments presented in this Rapid Communication we were constrained to a maximum usable power for the vertical dynamic tweezer of 1.2~W, set by the damage threshold of the conventional optical fiber carrying light to our 2D scanner. This bottleneck prevented us from extending the ghost evaporation scheme beyond two sites at this stage, despite disposing over 50~W of laser output power. We stress that this is purely a technical limitation which can be circumvented by using a free space link or a specialized high power photonic crystal fiber for the delivery. Hence the runaway scheme should be readily applicable to BEC production in multi-well configurations in excess of what is demonstrated here and what is achievable with trap weakening evaporation.

In conclusion, we have considered parallel BEC production within two settings of a multiplexed optical tweezer system. First, using standard trap weakening evaporation, four individual BECs were simultaneously attained, by lowering the power of a common horizontal guide beam. The condensates can be shuttled independently over several millimeters along the guide, providing an interesting platform for, e.g., sensing applications. Second, we provided an experimental demonstration of a ghost beam protocol for runaway evaporation and efficient BEC production in crossed optical dipole traps, superior to the standard scheme. The ghost laser beam controlling evaporation has a diameter comparable to that of the trap, and, unlike existing runaway schemes, affords flexible and targeted, \textit{individual} site addressing for multi-trap configurations.  As recently highlighted in \cite{Muessel2014a}, arrays of low-number BECs can offer distinct advantages over a single large condensate for quantum metrology applications.  Contrasting multi-BEC production in arrays set up by laser beam interference \cite{Hadzibabic2004,Muessel2014a}, a multiplexed tweezer approach offers individually addressed condensates, which can be moved on a centimetric scale. Hybrids of these two approaches would seem very appealing for applications such as quantum-enhanced magnetic gradiometry. We finally note that our approach straightforwardly extends to 2D arrays of crossed traps. This, for example, paves the way for interferometric generation of vortices \cite{Scherer2007} and vortex lattices \cite{private} from a multi-BEC source.
\Head{Acknowledgements}
We are grateful to Julia Fekete, Ana Rakonjac, and Kris Roberts for technical assistance, and we thank Bianca Sawyer for comments on our manuscript.
This work was supported the Marsden Fund of New Zealand (Contract No.~UOO1121) and a University of Otago Research Grant.

%

\end{document}